\documentstyle[preprint,prl,aps]{revtex}

\newenvironment{Fig}[1]{
\begin{figure}
\noindent\begin{minipage}[t]{\linewidth}
\begin{center}
\leavevmode \epsfxsize15.5cm\epsfysize20.0cm \epsfbox{#1} } {
\end{center}
\end{minipage}
\end{figure}
}
\def\q{{\bf{q}}}

\begin{document}
\draft
\input epsf
\title{Microscopic Theory of Damon-Eshbach Modes in Ferromagnetic Films}
\author{R. V. Leite
and R. N. Costa Filho}

\address{Departamento de F\'{\i}sica, Universidade Federal do Cear\'a,
Caixa Postal 6030\\ Campus do Pici, 60451-970 Fortaleza, Cear\'a,
Brazil\\E-mail: rai@fisica.ufc.br, FAX:+-55-85-288 9903}
\maketitle

\begin{abstract}
The surface spin wave branches in ferromagnetic films are studied
using a microscopic theory which considers both magnetic
dipole-dipole and Heisenberg exchange interactions. The dipole
terms are expressed in a Hamiltonian formalism, and the dipole
sums are calculated in a rapidly convergent form. The
Damon-Eshbach surface modes are analyzed for different directions
of the spin-wave propagation and also for different ratios of the
strength of the dipole interactions relative to the exchange
interactions. Numerical results are presented using parameters for
Fe and GdCl$_3$.

\end{abstract}
\vskip1truecm

\pacs{Keywords: Ferromagnetic Films, Spin Waves, Magnons, Dipolar
Interactions.}

\bigskip

It is well known theoretically that, in quasi-two-dimensional
systems, such as ultrathin ferromagnetic films, the short-range
exchange interactions alone are not necessarily sufficient to
establish a Ferromagnetic ordering above the ground state, and
that it is necessary to take into account the anisotropy and
long-range character of the dipolar interactions\cite{Mermin}.
Regarding the spin dynamics in ferromagnetic films, a macroscopic
(or continuous medium) theory for the dipole-dominated regime was
given by Damon and Eshbach\cite{Damon} in terms of magnetostatic
modes. They identified a surface branch of the spectrum, now known
as the Damon-Eshbach (DE) mode. Recently, microscopic theories
have been used to study the dipole-exchange spin waves (SW) in
ferromagnetic and antiferromagnetic
films\cite{Erickson,Camley,Costa}.

The DE mode has a number of striking and unusual properties. For
example, the existence condition of this mode depends on the
propagation angle relative to the principal axes of the crystal.
The x-direction in our case, being within the range
$0\ll\theta<\theta_c$, where $\theta_c$ depends on the
characteristics of the system, such as the applied field and
dipolar strength (see Ref. 6 for a review). The aim of this paper
is to study the DE modes in ferromagnetic films at low
temperatures ($T\ll T_c$, where $T_c$ is the Curie temperature) by
using a microscopic theory, and including both magnetic
dipole-dipole and Heisenberg exchange interactions. The dipole
terms are expressed in a Hamiltonian formalism, and the dipole
sums are calculated in a rapidly convergent form. In order to find
the critical $\theta_c$ where the DE modes vanish, we have to
study the dependence of the DE modes and the SW energies (for each
of the discrete branches) on the variation of the exchange
parameter, on the strength of the dipolar interaction, and on the
wave vector direction of propagation.  Another important factor in
the analisys is  the effect of different dipole interaction
strengths (relative to the exchange)on the dispersion relation of
the SW. Numerical results are presented using parameters for Fe
and GdCl$_3$.

Consider a ferromagnetic film with $N$ atomic layers arranged on a
simple cubic lattice, with lattice constant $d$. The wave vector
$\q$ makes an angle $\theta$ with respect to the $x$-axis, and the
Zeeman field $H_0$ is assumed to be parallel to $z$ direction. The
surface of the film is in the $x-z$ plane, while the $y$-axis is
perpendicular to the film. The Hamiltonian can be written as
\begin{equation}
H=-\frac{1}{2}\sum_{ij}J_{ij}{\bf S}_i\cdot{\bf S}_j-g\mu_BH_0
\sum_iS^z_i+\frac{1}{2}(g\mu_B)^2\sum_{ij}D^{\alpha\beta}_{ij}
S^\alpha_iS^\beta_j,
\end{equation}
where ${\bf S}_i$ is the spin operator at site $i$ and $J_{ij}$ is
the exchange coupling between sites labeled $i$ and $j$. We assume
that the exchange coupling is $J_1$ and $J_2$ for nearest and
next-nearest neighbors respectively. The second term of the above
equation is the Zeeman term, and the dipole-dipole interaction
between spins is represented by the last term, where $\alpha$ and
$\beta$ denote the components $x,y,$ or $z$, and
\begin{equation}
D^{\alpha\beta}_{ij}=\{|{\bf
r}_{ij}|^2\delta_{\alpha\beta}-3r^\alpha_{ij}
r^\beta_{ij}\}/{|{\bf r}_{ij}|} ^5
\end{equation}
whit ${\bf r}_{ij}={\bf r}_{j}-{\bf r}_{i}$, and the case $i=j$
excluded from the sums in Eq. (1).

Using the Holstein-Primakoff transformations, the Hamiltonian can
be expanded as $H=H^{(2)}+H^{(3)}+H^{(4)}+\ldots$ (apart from a
constant), where $H^{(m)}$ denotes the term involving a product of
$m$ boson operators. Here we are considering the non-interacting
(linear) SW modes that are described by the quadratic term
$H^{(2)}$. Therefore, the Hamiltonian takes the form:
\begin{equation}
H=\sum_{qnn'}\{A_{nn'}(\q)a^\dag_{qn}a_{qn'}+
B_{nn'}(\q)[a_{qn}a_{-qn'}+a^\dag_{qn}a^\dag_{-qn'}]\}.
\end{equation}
Here we are  using a representation of the boson operators
$a^\dag$ and $a$ in terms of a $2D$ wave vector
$\q=(q\cos{\theta},q\sin{\theta})$, q being the modulus of $\q$,
parallel to the film surface and indices $n$, $n'$ $(=1,2,...,N)$
that label the atomic layers parallel to the surface. The
coefficients $A_{nn'}(\q)$ and $B_{nn'}(\q)$ are described in
Ref.5. A linear transformation to new boson operators can then be
found such that $H$ takes a diagonalized form \cite{Kontos}. Its
eingenvalues yield the energies $E_\nu(\q)$ of the discrete SW
branches, where $\nu$(=1,2,...,N) labels the branches.

In Fig. 1(a), we show the behavior of the lowest SW branches for
very small wave vectors in a GdCl$_3$ film with $N=20$. For this
material, we use the parameters; $4\pi M= (g\mu_B S)/d^3 = 0.82$ T
to characterize the dipolar strength, the bulk exchange field
$H_{Ex} =6S(J_1+2J_2)= 0.54$ T, and $H_0= 0.36$ T. We have taken
the case of $\theta=0$ (the Voigt geometry), for which it is known
that there is a Damon-Eshbach (DE) surface mode\cite{Damon} in the
magnetostatic continuum limit where exchange is neglected. It can
be seen that a purely magnetostatic DE mode starts at $q_x=q=0$ at
the frequency $18.25$ GHz. There are hybridizations until the
eighth mode, and the DE mode tends to be flat around the value
$21.5$ GHz with increasing $q_x$. On the other hand, in a material
such as Fe that has a much smaller value for the ratio $4\pi M /
H_{Ex}$, it is necessary to consider a much larger film thickness
in order for the DE mode to hybridize with the lowest bulk branch.
This is because in a thicker film, the density of modes increases
and create more modes in the region where the DE mode propagates,
making hybridizations possible. An example for Fe is given in Fig.
1(b) considering $N = 120$, $4\pi M = 2.14$ T, $H_{Ex} = 2140$ T,
and $H_0= 0.054$ T. Only the lowest branches of the spectrum are
shown in Fig. 1(b), where it can be seen that much smaller values
of $q_x$ are necessary (compared with the previous example) to see
the analogue of the DE mode in the spectrum.

In Fig. 2, we show results for the SW frequencies of the
above-mentioned materials considering the SW propagation direction
$\theta$. In Fig. 2(a), the dependence of the lowest GdCl$_3$ SW
branches on $\theta$ is shown for a fixed value $qa/\pi$. For
small wavevectors, the branches decrease their frequency values as
$\theta=0$  goes from 0 to $\theta=\pi/2$. As in Fig. 1, the
hybridization of the modes is seen to decrease until the second
mode. The two lowest branches are the most affected by the SW
propagation direction. The others tend to become flat after
certain angle values. The figure also shows the reciprocity of the
dispersion relation as the angle goes to $\pi$. The same effect
can be observed in Fig. 2(b) considering Fe. In this case, the
dependence of the modes  on the angle is much weaker than in the
case of GdCl$_3$, and only the two lowest  modes are affected by
the propagation angle.

In conclusion, we have used a microscopic theory to study
Damon-Eshbach SW modes in the dipole-exchange regime in thin
ferromagnetic films.  We analyzed the dependence of the DE modes
on the SW propagation direction and the ratio between dipolar and
exchange strengths. The results show that the DE modes tends to
vanish as the system change from the Voigt geometry to a general
propagation direction, and that this change depends on the
strength of the ratio between the dipole and exchange
interactions.

We acknowledge the financial support from the FUNCAP, CAPES and
CNPq, Brazilian agencies.

\newpage

\centerline{\bf FIGURE CAPTIONS}

\bigskip

Fig.1 The linear SW dispersion relation for the lowest branches of
(a) a 20-layer GdCl$_3$ film and (b) a 120-layer Fe film. In both
cases $\theta=0$.

\bigskip

Fig.2 The SW frequencies as a function of the propagation
direction $\theta$ for (a)  a 20-layer GdCl$_3$ film taking
$qa/\pi=0.02$ and (b) a 120-layer Fe film for $qa/\pi=0.002$.

\newpage

\begin{Fig}{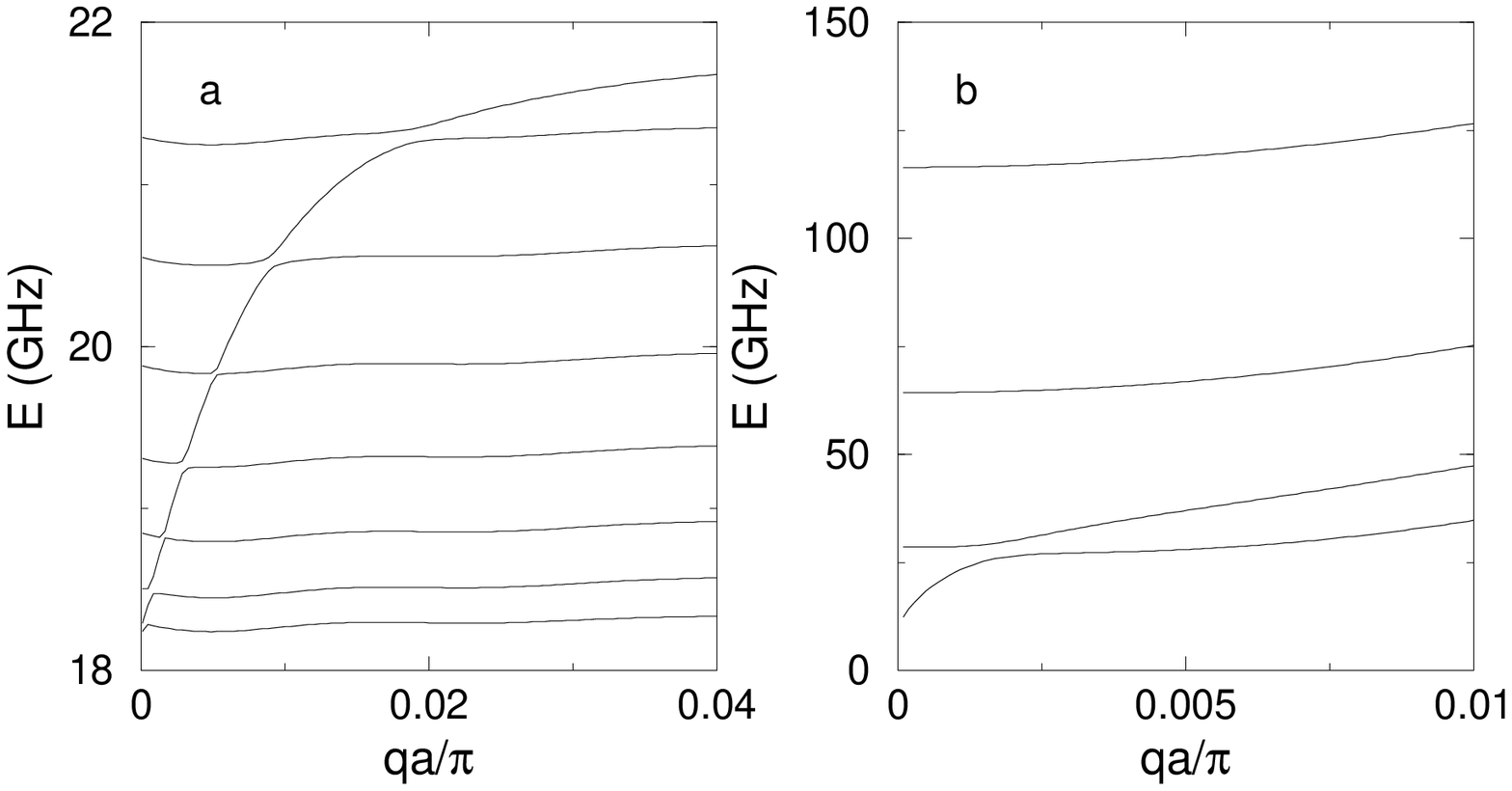}
\caption{R.V. Leite at al.}
\end{Fig}
\vskip-2.25cm

\newpage

\begin{Fig}{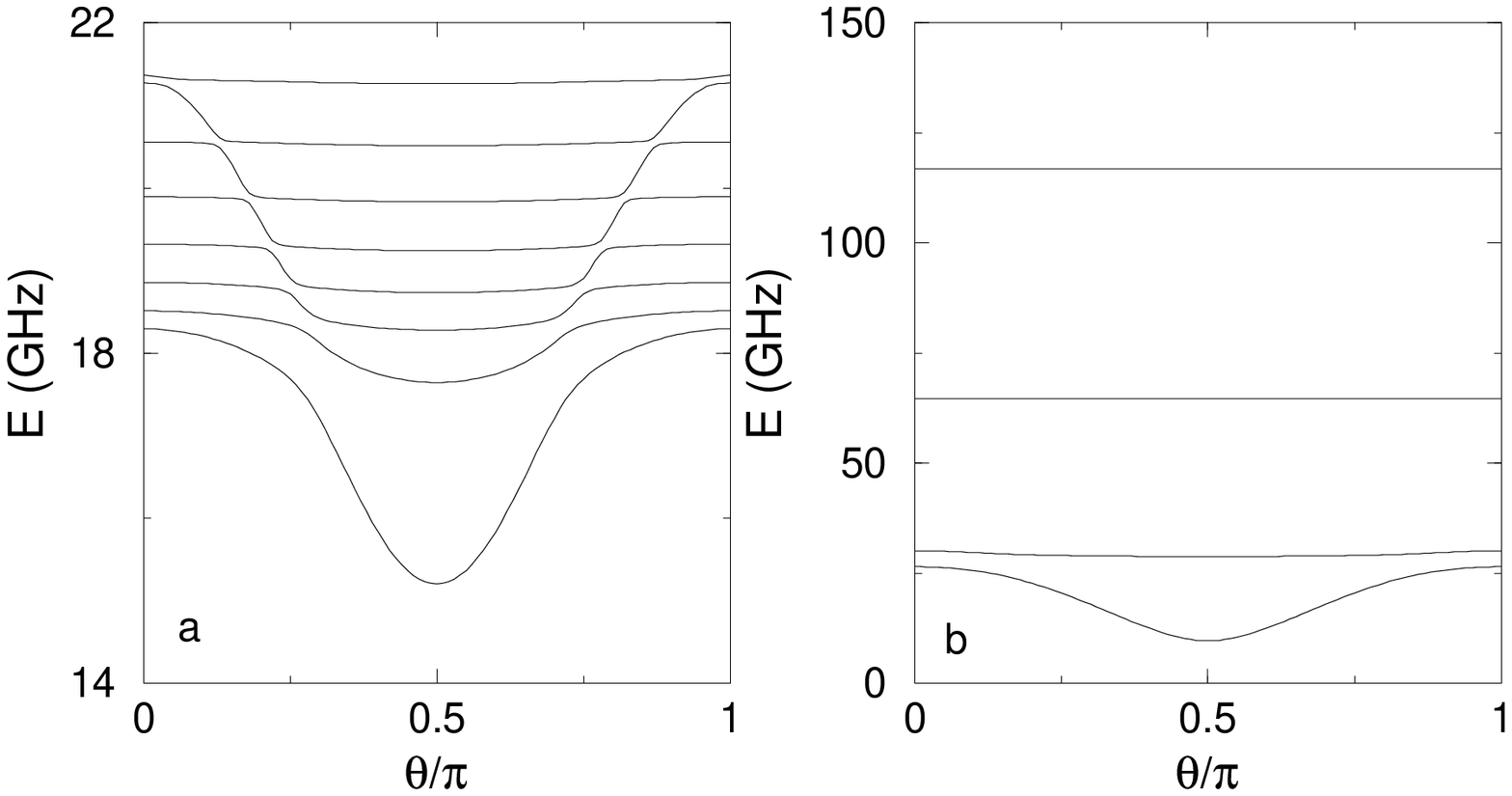}
\caption{ R.V. Leite at al.}
\end{Fig}

\end{document}